\documentclass[letter,oneside,9pt,twocolumn,article]{aiaa}
\usepackage{graphicx}
\usepackage{rotating}
\usepackage{bm}
\usepackage{flushend}
\usepackage{cite}
\usepackage[super,comma]{natbib}
\usepackage[bf,footnotesize]{subfigure}
\usepackage{adjustbox}
\usepackage{tabularx}
\usepackage{multirow}
\usepackage{multicol}
\usepackage{indentfirst} 
\usepackage[hidelinks]{hyperref}
\usepackage{float}
\usepackage{xcolor}

\usepackage{mwe}
\usepackage[markcase=noupper]{scrlayer-scrpage}
\usepackage{caption} 

\ihead*{}
\ohead*{\pagemark}
\ifoot*{\rm Submitted to Physics of Fluids}
\ofoot*{\rm Unrestricted content}
\newcommand{\mfd}{\displaystyle}

\author{
Bernard Parent\thanks{Associate Professor, bparent@arizona.edu.} ~~and~ Felipe Martin Rodriguez Fuentes\thanks{Graduate Student}\\[0.3em] \it University of Arizona, Tucson, AZ 85721, USA.
}

\title{
Thermodynamically Consistent Vibrational-Electron Heating: Generalized Derivation for Excited State Populations
}

\abstract{ 
Accurate prediction of electron temperature ($T_{\rm e}$) in non-equilibrium plasma flows is critical for applications ranging from hypersonic flight to plasma-assisted combustion. We recently proposed a thermodynamically consistent model for vibrational-electron (V-e) heating [Phys. Fluids 37, 096141 (2025)] which enforces convergence of $T_{\rm e}$ to the vibrational temperature ($T_{\rm v}$) at equilibrium. While the original derivation assumed electron energy loss was dominated by collisions with ground-state molecules, this Letter presents a rigorous generalization of the model. We demonstrate that the heating-to-cooling ratio $\exp(\theta_{\rm v}/T_{\rm e}-\theta_{\rm v}/T_{\rm v})$ with $\theta_{\rm v}$ the characteristic vibrational temperature remains valid even when electron cooling interactions with vibrationally excited states are included. This derivation removes the previous constraint assuming ground-state dominance, thereby extending the model's validity to plasma flows  where vibrationally excited populations contribute significantly to electron cooling.  
}

\nomenclature{
\begin{nomenclaturelist}{Roman symbols}

\item[$C_k$]{charge of $k$th  species, C}
  \end{nomenclaturelist}

  \begin{nomenclaturelist}{Greek symbols}
\item[$\beta_{k}^{\rm n}$]{parameter equal to 1 when $k$th species is neutral, 0 otherwise}
  \end{nomenclaturelist}

}

\begin{document}
\maketitle

The energy exchange between free electrons and molecular vibrational modes, known as electron-vibrational (e-V) coupling, is a cornerstone process in numerous applications involving non-equilibrium plasmas. Its accurate representation is vital for modeling hypersonic flight, plasma-assisted combustion (PAC), and laser-induced plasmas (LIP).

In hypersonic flight beyond Mach 10, the plasma layer enveloping a vehicle critically influences its operation. This plasma enables advanced technologies such as electromagnetic shielding, as explored in early work by Musal\cite{gmrl:1963:musal} and later by Gregorie, \cite{dtic:1992:gregorie} as well as plasma antennas, \cite{ieee:2024:magarotto} electron transpiration cooling, \cite{aip:2017:hanquist,aiaapaper:2021:parent} and magnetohydrodynamic (MHD) systems. \cite{aiaaconf:2022:moses,jtht:2023:parent,aiaa:2025:parent} The viability of these technologies hinges on precise predictions of plasma properties like electrical conductivity and plasma frequency, which are strongly dependent on the electron temperature ($T_{\rm e}$). Indeed, the collision cross sections needed to determine electrical conductivity, for instance, are a function of electron temperature. Furthermore, electrical conductivity depends on the plasma density, which is also a function of $T_{\rm e}$ through processes like electron-ion recombination and ambipolar diffusion to surfaces. \cite{pf:2022:parent} Plasma density, in turn, governs the
plasma frequency, which is crucial for electromagnetic interactions. In the characteristic non-equilibrium of re-entry flows, where nitrogen is not fully dissociated, e-V coupling is the dominant mechanism governing $T_{\rm e}$. Specifically, superelastic collisions with vibrationally excited N$_2$ molecules become the primary heating source for electrons, yet direct data for these heating rates are scarce.

Similarly, e-V coupling is central to PAC. High-energy electrons create radicals and excited species while also initiating thermal pathways. These pathways include fast heating from the quenching of electronic states \cite{apl:2011:bak} and slow heating from the vibrational-translational (V-T) relaxation of N$_2$ molecules excited via e-V transfer. Detailed kinetic mechanisms \cite{rsta:2015:adamovich,jap:2023:chen} model this process, whereas phenomenological approaches often assume a fixed energy partition. \cite{cf:2016:castela} However, recent studies have shown this energy partition is not fixed and varies with operating conditions, \cite{cf:2025:dijoud} necessitating a predictive model for e-V energy transfer to accurately simulate PAC.

Laser-induced plasmas also require accurate e-V coupling models. \cite{jap:2019:peters, jap:2010:shneider, jap:2023:pokharel} In these systems, electron-impact vibrational excitation is a major energy loss for electrons, while the reverse process—heating via superelastic collisions—critically slows the decay of $T_{\rm e}$ over time. This is paramount because $T_{\rm e}$ controls the rates of electron attachment and recombination, which dictate the plasma's lifetime.

Modeling the vibrational-electron heating rate has been a persistent challenge. Early phenomenological models scaled the heating rate with simple temperature or energy ratios. \cite{jap:2010:shneider,jtht:2012:kim,jtht:2013:farbar} Although these drive the system towards equilibrium ($T_{\rm e} = T_{\rm v}$), they lack a rigorous foundation. More recent models applied the principle of detailed balance, \cite{jap:2019:peters,jap:2023:pokharel} but their formulation contained a crucial flaw that prevents the convergence of electron and vibrational temperatures at equilibrium, violating thermodynamic consistency.

To address these shortcomings, we recently developed a novel, thermodynamically consistent model for vibrational-electron heating in Ref.~\citenum{pof:2025:rodriguez}. That work established a robust heating-to-cooling ratio derived rigorously from detailed balance by assuming a Boltzmann distribution for vibrational states. However, the original derivation relied on the simplifying assumption that the macroscopic electron cooling rate was dominated by collisions with molecules in the ground state. In this Letter, we extend that proof to include electron cooling by vibrationally excited states. This new proof extends the applicability of the referenced model but does not change its recommended formula.

We derive this generalization from basic principles, ensuring the model satisfies the detailed balance principle and remains thermodynamically consistent. We begin by defining an effective activation energy as the average activation energy of all vibrationally excited states:
\begin{equation}
 \mathcal{E}_{\rm eff}
\equiv
\frac{\textrm{vibrational energy excluding zero-point energy}}{\textrm{number of vibrationally excited molecules}}
\end{equation}

Here, $e_{\rm v}$ denotes the mass-specific vibrational energy, exclusive of zero-point energy. By designating $N_{v}$ as the number density of the molecules occupying the $v$th vibrationally excited level, we express the relationship as:
\begin{equation}
\mathcal{E}_{\rm eff}=\frac{\rho e_{\rm v}}{\mfd\sum_{v=1}^\infty N_{v}}
\label{eqn:eff_act_energy}
\end{equation}
Replacing the mass density $\rho$ with the product of the total number density $N$ and the molecular mass $m$—where $N$ explicitly encompasses the populations of both the ground state and all excited states—allows for a reformulation in terms of vibrational fractions. Letting $f_v$ represent the population fraction of the $v$th vibrational level, Eq.~(\ref{eqn:eff_act_energy}) transforms into:
\begin{equation}
\mathcal{E}_{\rm eff} \sum_{v=1}^\infty f_v = m  e_{\rm v}
\label{eqn:eff_act_energy_3}
\end{equation}
Applying the normalization condition where the summation of fractions over all levels (including the ground state $v=0$) equates to unity, we arrive at:
\begin{equation}
\left(1- f_{v=0} \right) \mathcal{E}_{\rm eff} = m e_{\rm v}
\label{eqn:eff_act_energy_4}
\end{equation}
Following  Ref.~\citenum{book:1989:anderson} (pp.~518, 578), we can obtain $f_v$ from the Boltzmann distribution given $T_{\rm v}$:
\begin{equation}
f_v = 
\exp\left(\frac{-\mathcal{E}_v }{k_{\rm B} T_{\rm v}}\right)
\left/
\sum_{n=0}^{\infty} \exp\left(\frac{-\mathcal{E}_{n}}{ k_{\rm B} T_{\rm v}}\right) 
\right.
\label{eqn:vib_frac_full}
\end{equation}
Here, $\mathcal{E}_{v}$ denotes the energy of the $v$th vibrational level. The invocation of an equilibrium distribution is justified provided that vibrational-vibrational and vibrational-translational relaxation mechanisms are sufficiently rapid. Furthermore, we adopt the harmonic oscillator approximation yielding an activation energy of the $n$th state equal to $\mathcal{E}_n= \theta_{\rm v} k_{\rm B} n$. In this context, $\theta_{\rm v}$ represents the characteristic vibrational temperature. Consequently, the ground state fraction becomes:
\begin{equation}
f_{v=0}=\frac{1}{\mfd \sum_{n=0}^\infty \exp\left(-\frac{n \theta_{\rm v}}{T_{\rm v}}\right)} 
\end{equation}
It can be shown that the infinite geometric series on the denominator becomes $(1-\exp(-\theta_{\rm v}/T_{\rm v}))^{-1}$. Then, the fraction of the ground state simplifies to:
\begin{equation}
f_{v=0} = 1 - \exp({-\theta_{\rm v} / T_{\rm v}}) 
\label{eqn:vib_frac_harmonic}
\end{equation}
The effective activation energy can then be simplified by substituting Eq.~(\ref{eqn:vib_frac_harmonic}) in Eq.~(\ref{eqn:eff_act_energy_4}):
\begin{equation}
\mathcal{E}_{\rm eff}
=
\frac{m e_{\rm v}}{\exp(- \theta_{\rm v} / T_{\rm v})}
\end{equation}
Using the Planck distribution for a harmonic oscillator, we obtain the average vibrational energy $e_{\rm v}$ net of zero-point energy:
\begin{equation}
e_{\rm v} = \dfrac{R \theta_{\rm v}}{\exp(\theta_{\rm v}/T_{\rm v})-1}
\label{eqn:ev_Tv}
\end{equation}
where $R$ is the specific gas constant of the molecule under consideration. Substituting the latter equation in the former and noting that $R=k_{\rm B}$, the effective activation energy simplifies to:
\begin{equation}
\mathcal{E}_{\rm eff}
= \dfrac{k_{\rm B}\theta_{\rm v}}{1-\exp(-\theta_{\rm v}/T_{\rm v})}
\label{eqn:Eeff_Tv}
\end{equation}
Temporarily setting the latter aside, the second step of our derivation defines an \textit{effective} electron cooling reaction rate, $k_{\rm e-v}$. This rate is constructed such that, when multiplied by the effective activation energy, it recovers the electron-vibrational energy loss caused by inelastic collisions. Thus, the effective rate is given by:
\begin{equation}
  k_{\rm e-v}   
\equiv \frac{1}{N \mathcal{E}_{\rm eff}} \sum_{v=0}^\infty \sum_{l} N_{v} k_{vl} \mathcal{E}_{vl}
\label{eqn:kcool}
\end{equation}
Here, $N_v$ denotes the number density of molecules in the $v$-th vibrationally excited state, where $v=0$ corresponds to the ground state. The terms $\mathcal{E}_{vl}$ and $k_{vl}$ represent the activation energy and reaction rate, respectively, for the $l$-th electron-vibrational cooling process involving state $v$.

While the principle of detailed balance applies strictly at the microscopic level, the macroscopic forward (cooling) and reverse (heating) rates are obtained by integrating over the same electron temperature distribution. Consequently, the resulting equilibrium constant satisfies the detailed balance relation, allowing us to express the effective heating reaction rate as follows:
\begin{equation}
k_{\rm v-e}=k_{\rm e-v} \exp \left(\frac{\theta_{\rm e}}{T_{\rm e}} \right)
\label{eqn:kheat}
\end{equation}
where $\theta_{\rm e}$ is a closure coefficient proportional to the activation energy which will be determined later.  

Since inelastic electron heating results from the collisional de-excitation of vibrational states, it can be written as the product of the effective heating rate $k_{\rm v-e}$, the average activation energy, and the populations of the vibrationally excited molecules and electrons:
\begin{equation}
Q_{\rm v-e}=N_{\rm e} \sum_{v=1}^\infty N_{v} k_{\rm v-e} \mathcal{E}_{\rm eff}
\end{equation}
But recall that the definition of the effective activation energy outlined in Eq.~(\ref{eqn:eff_act_energy}) entails that $\sum_{v=1}^\infty N_{v} \mathcal{E}_{\rm eff}=\rho e_{\rm v}$. Further, we can substitute $k_{\rm v-e}$ from Eq.~(\ref{eqn:kheat}) to obtain:
\begin{equation}
Q_{\rm v-e}=N_{\rm e} \rho e_{\rm v} k_{\rm e-v} \exp \left(\frac{\theta_{\rm e}}{T_{\rm e}} \right)
\end{equation}
Further substituting $k_{\rm e-v}$ from Eq.~(\ref{eqn:kcool}), it follows that:
\begin{equation}
Q_{\rm v-e}=N_{\rm e} \rho  \frac{e_{\rm v}}{\mathcal{E}_{\rm eff} N } 
\exp \left(\frac{\theta_{\rm e}}{T_{\rm e}} \right)
\sum_{v=0}^\infty \sum_{l} N_{v}k_{vl} \mathcal{E}_{vl} 
\end{equation}
Finally, we can rewrite $\mathcal{E}_{\rm eff}$ and $e_{\rm v}$ in terms of $T_{\rm v}$ using Eqs. (\ref{eqn:ev_Tv}) and (\ref{eqn:Eeff_Tv}) to arrive at:
\begin{equation}
Q_{\rm v-e}=N_{\rm e} \frac{\rho}{N} \dfrac{R}{k_{\rm B}}
\dfrac{1-\exp(-\theta_{\rm v}/T_{\rm v})}{\exp(\theta_{\rm v}/T_{\rm v})-1} 
\exp \left(\frac{\theta_{\rm e}}{T_{\rm e}} \right)
\sum_{v=0}^\infty \sum_{l} N_{v}k_{vl} \mathcal{E}_{vl}
\end{equation}
We multiply both the numerator and denominator by $\exp(\theta_{\rm v}/T_{\rm v})$. Using the relations $\rho=N m$ and $R=k_{\rm B}/m$, and upon simplification, we obtain:
\begin{align}
Q_{\rm v-e}
&= 
 \exp \left(\dfrac{\theta_{\rm e}}{T_{\rm e}}-\dfrac{\theta_{\rm v}}{T_{\rm v}}\right) 
 \underbrace{ N_{\rm e} \sum_{v=0}^\infty N_{v } \sum_{l} k_{vl} \mathcal{E}_{vl} }_{\rm rate~of~e-V~cooling}
\label{eqn:Qe_inelastic_heating}
\end{align}
Noting that the last terms on the RHS correspond to the rate of {electron-vibrational} cooling by the molecule, we can say:
\begin{align}
\frac{Q_{\rm v-e}}{Q_{\rm e-v}}
&= 
 \exp \left(\dfrac{\theta_{\rm e}}{T_{\rm e}}-\dfrac{\theta_{\rm v}}{T_{\rm v}}\right) 
\label{eqn:scale_proposed_preliminary}
\end{align}
To determine the closure coefficient $\theta_{\rm e}$, we impose a condition of thermodynamic consistency: the ratio of inelastic electron heating to cooling must be unity when the electron temperature equals the vibrational temperature ($T_e = T_v$). This requirement ensures vanishing net energy exchange in the limit of thermal equilibrium. It follows that this condition is satisfied when:
\begin{equation}
\theta_{\rm e}=\theta_{\rm v}
\end{equation}
Recalling Eq.~(\ref{eqn:kheat}) and defining $\theta_{\rm v}=\mathcal{E}_{v=1}/k_{\rm B}$ (where $\mathcal{E}_{v=1}$ is the activation energy of the first vibrational level), the choice of $\theta_{\rm e}=\theta_{\rm v}$ ensures thermodynamic consistency by anchoring the equilibrium constant to the fundamental transition ($v=0 \to 1$). Physically, this implies that the system is dominated by single-quantum transitions, effectively neglecting overtones. 

Consequently, although the approximation $\theta_{\rm e}=\theta_{\rm v}$ does not strictly satisfy detailed balance, it yields a solution that closely mimics the principle while guaranteeing macroscopic thermodynamic consistency. On this basis, we propose the following electron heating model:
\begin{align}
\frac{Q_{\rm v-e}}{Q_{\rm e-v}}
&= 
 \exp \left(\dfrac{\theta_{\rm v}}{T_{\rm e}}-\dfrac{\theta_{\rm v}}{T_{\rm v}}\right) 
\label{eqn:scale_proposed}
\end{align}

The latter model makes two important assumptions. It assumes a Boltzmann distribution for the vibrationally excited states, and no overtones during the electron cooling process. As explained in Ref.~\citenum{pf:2025:rodriguez:2}, such is well justified for many planetary entry flows.  While the generalized derivation holds for fundamental transitions, we acknowledge that in high-energy regimes ($T_{\rm e} > 1$ eV), such as during the pulse phase of PAC or LIP, overtone processes become non-negligible. In these specific high-electron-temperature windows, a detailed kinetic approach accounting for level-skipping\cite{jap:2019:peters} is preferable. However, as the plasma cools in the afterglow and $T_e$ drops below 1~eV---critical for determining recombination rates and plasma lifetime---the flow enters the regime where the fundamental transition dominates. Here, our thermodynamically consistent formulation should be employed to ensure the physically correct convergence of $T_{\rm e}$ to $T_{\rm v}$, preventing the non-physical divergence often observed with other formulations near equilibrium.

In conclusion, we have extended the domain of validity for the thermodynamically consistent electron heating model. By proving that the macroscopic heating rate formulation remains invariant regardless of the population of excited states (assuming a Boltzmann distribution), we establish the model’s robustness for plasma flows where a significant portion of the electron cooling takes place with vibrationally-excited states. Unlike previous state-to-state approaches which failed to yield zero net energy exchange at thermal equilibrium, this generalized derivation mathematically enforces the correct convergence of electron temperature to vibrational temperature. Consequently, the model ensures the correct physical behavior in post-shock and relaxation flows encountered in aerospace applications.

\section{Data Availability}

Data sharing is not applicable to this article as no new data were created or analyzed in this study.

\footnotesize
\bibliography{all}
\bibliographystyle{aiaa2}
\end{document}